\documentclass{aipproc}

\layoutstyle{6x9}

\newcommand\doingARLO[2][]{%
  \ifx\mmref\undefined #1\else #2\fi
}

\def\tobs{t_{\rm obs}}

\def\Rdec{R_{\rm dec}}
\def\Rgap{R_{\rm gap}}
\def\Racc{R_{\rm acc}}
\def\Rload{R_{\rm load}}

\def\Rtrail{R_{\rm trail}}
\def\Rb{R_\beta}

\def\Gn{\Gamma_n}

\def\Gej{\Gamma_{\rm ej}}

\def\be{\begin{equation}}
\def\ee{\end{equation}}

\newbox\grsign \setbox\grsign=\hbox{$>$} \newdimen\grdimen \grdimen=\ht\grsign
\newbox\simlessbox \newbox\simgreatbox \newbox\simpropbox
\setbox\simgreatbox=\hbox{\raise.5ex\hbox{$>$}\llap
     {\lower.5ex\hbox{$\sim$}}}\ht1=\grdimen\dp1=0pt
\setbox\simlessbox=\hbox{\raise.5ex\hbox{$<$}\llap
     {\lower.5ex\hbox{$\sim$}}}\ht2=\grdimen\dp2=0pt
\setbox\simpropbox=\hbox{\raise.5ex\hbox{$\propto$}\llap
     {\lower.5ex\hbox{$\sim$}}}\ht2=\grdimen\dp2=0pt

\begin{document}

\title 
     [Early Stages of the GRB Explosion]
      {Early Stages of the GRB Explosion}

\classification{43.35.Ei, 78.60.Mq}
\keywords{Document processing, Class file writing, \LaTeXe{}}

\author{A. M. Beloborodov}{
  address={Physics Department, Columbia University, 538  West 120th
           Street New York, NY 10027},
  email={amb@phys.columbia.edu}
}



\begin{abstract}
Physics of GRB blast waves is discussed with a focus on two effects: 
(1) pair creation in the external medium by the gamma-ray front and 
(2) decay of neutrons ahead of the decelerating blast wave. 
Both effects impact the afterglow mechanism at radii up to $10^{17}$~cm.
\end{abstract}

\date{\today}

\maketitle

\section{Introduction}

GRB afterglow is well explained as emission from a decelerating relativistic
blast wave. 
Most of the afterglow data collected to date are obtained relatively late, 
hours or days after the prompt GRB, when 
the blast wave is already at final stages of deceleration.
{\it Swift} satellite will provide the missing data on the early stage
when the afterglow is most luminous. 

The blast wave deceleration begins
at a radius $\Rdec\sim 10^{15}-10^{17}$~cm, which depends
on the ambient density $n_0$ and the initial Lorentz factor $\Gamma_0$ of
the blast wave.\footnote{$\Gamma_0$ can be smaller than the ejecta Lorentz 
factor $\Gej$ if the reverse shock in the ejecta is relativistic, which is 
the case for a dense external medium.}
$\Rdec$ does not exceed $10^{17}$~cm, the estimated
fireball size during the late afterglow, and can be one or two orders
smaller than $10^{17}$~cm, especially if the circumburst medium has
$R^{-2}$ density profile.
Three effects are predicted to occur at $R\sim 10^{16}$~cm:

\begin{itemize}

\item
The reverse shock crosses the ejecta and can produce a detectable flash
of soft emission (if the ejecta are not dominated by the Poynting flux)
\cite{MR93,SP99}.

\item
The external medium ahead of the forward shock is 
loaded with a large number of $e^\pm$ pairs by the 
prompt $\gamma$-ray front \cite{TM00,MRR01,B02}.
The leptonic component of the preshock medium is enriched by 
$e^\pm$, which leads to a dramatic softening of the early afterglow.

\item
The neutron component of the GRB ejecta overtakes
the decelerating blast wave and deposits energy and momentum into the 
external medium by $\beta$-decay \cite{B03b}. 
The leading neutron front leaves behind a relativistic trail
--- a hot mixture of the decay products and ambient particles.
The forward shock of the blast wave propagates
in this trail instead of the customary ambient medium. 
The impact of neutrons lasts about 10 e-foldings of the $\beta$-decay,
and at $\Rtrail\approx 10^{17}$cm the trail becomes static and cold, i.e. 
indistinguishable from a normal ambient medium.

\end{itemize}

The paradox of relativistic explosions
is that even a small fraction of their energy deposited by a precursor
into the external medium impacts the ensuing blast wave and
the afterglow radiation. The importance of both 
$\gamma$-ray and neutron precursors is entirely due to the high 
Lorentz factor of the explosion. 
The study of both precursors together is a complicated physical problem
and we discuss them separately here.


\section{Electron-positron loading}
 
A medium overtaken by a front of collimated $\gamma$-rays is inevitably 
$e^\pm$-loaded \cite{TM00,MRR01,B02}. This happens because some $\gamma$-rays 
Compton scatter off the medium and get absorbed by the
primary collimated radiation via reaction $\gamma+\gamma\rightarrow e^++e^-$. 
The medium is optically thin, so only a tiny fraction of the prompt GRB 
scatters, however, the number of scattered $\gamma$-rays and created $e^\pm$
{\it per ambient electron} is huge, $n_\pm/n_0\gg 1$ . 

The pair loading factor $Z=1+n_\pm/n_0$ does not depend on the
ambient density 
and can be calculated starting with just one ambient electron.
The number of $\gamma$-rays scattered by the electron at a radius $R$
is proportional to the column density of the $\gamma$-ray front at this 
radius, $E_\gamma/4\pi R^2\propto R^{-2}$, and therefore $Z$ is high at 
small $R$. $Z(R)$ is determined by the isotropic equivalent of the prompt
GRB energy $E_\gamma$ (and slightly depends on the exact spectral shape 
of the prompt GRB). For the brightest bursts $E_\gamma>10^{54}$~erg, 
and a typical $E_\gamma\sim 10^{53}$~erg. 

The created $e^\pm$ also scatter radiation, which can lead to an
exponential runaway of pair creation. This runway takes place at radii 
\be
 R<\Rload\approx 2 \times 10^{16}(E_\gamma/10^{53})^{1/2}{\rm ~cm},
\ee
leading to $Z\gg 1$ \cite{B02}. The calculation shows also that
at $R<\Racc=\Rload/2.3$ the $e^\pm$-loaded medium is pushed to 
relativistic velocities $\beta\gamma>1$ by the $\gamma$-ray front.
The scattered fraction $\delta E_\gamma$ of the GRB radiation is proportional 
to the optical depth of the swept-up ambient mass $m$, which is proportional 
to $n_0$. However, the medium acceleration $\gamma=\delta E_\gamma/m c^2$ 
does not depend on $n_0$ ($m$ cancels out), so only $E_\gamma$ determines 
$\Racc$.

The pair-loading factor $Z(R)$ and the Lorentz factor $\gamma(R)$ 
of a medium overtaken by the GRB radiation front are shown in Fig.~1.
These parameters were calculated numerically and approximated by
analytical formulae in \cite{B02}. $Z$ varies exponentially between $\Racc$ 
and $\Rload=2.3\Racc$ and $Z(\Racc)\approx 74$. 
At $R<\Racc$ both $\gamma$ and $Z$ vary as power-laws with radius. 
An interesting phenomenon takes place at small radii $R<\Rgap\approx\Racc/3$: 
here the external shock may not exist at all because
the medium gains so high $\gamma$ that it runs away from the ejecta and a 
gap is opened.

\begin{figure}
  \includegraphics[height=.4\textheight]{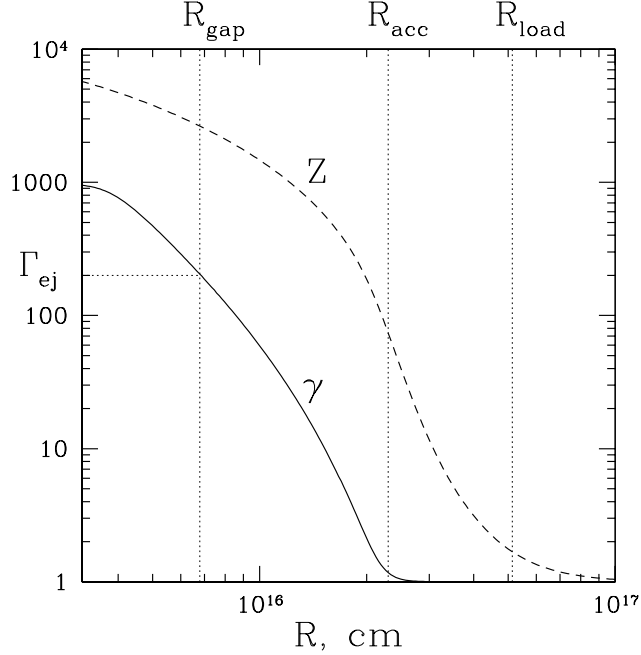}
  \caption{Pair-loading factor $Z=1+n_\pm/n_0$ and acceleration $\gamma$ 
of the external medium by the $\gamma$-ray front. As the front propagates
to larger radii $R$, the $e^\pm$ loading and acceleration effects are
reduced and become negligible at $R\approx 10^{17}$~cm.
The figure shows the results for GRBs with isotropic energy 
$E_\gamma=10^{54}$~erg. The corresponding curves for different 
$E_\gamma$ are obtained by re-scaling radius $R\rightarrow
R(E_\gamma/10^{54})^{1/2}$. 
Example ejecta Lorentz factor $\Gej=200$ is indicated; $\gamma=\Gej$
defines the gap radius $\Rgap$.
}
\end{figure}

The main observational effect of $e^\pm$-loading is a strong softening
of synchrotron emission from the forward shock. 
Indeed, the shock energy per proton, $\Gamma m_pc^2$, can now be shared by 
$Z$ leptons, and the energy per lepton is reduced by the factor $Z^{-1}\ll 1$. 
Then the frequency of 
synchrotron radiation $\nu_s$ is reduced as $Z^{-2}$. Accurate calculation 
shows that the preacceleration $\gamma$ also softens the emission because 
it reduces the pressure in the blast wave; as a result
$\nu_s\propto Z^{-2}\gamma^{-5/2}$. This has a strong effect on
the early afterglow: it starts as a very soft signal and later evolves to 
the normal X-ray emission.

The energy dissipated in the forward shock at $R<\Rload$ depends on 
$\Rdec/\Rload$ and can vary.
For example, in a medium with $n_0\sim 1-10$~cm$^{-3}$ 
(standard ISM) it can be estimated as
$E_{\rm soft}\approx E (\Rload/\Rdec)^3=(10^{-3}-10^{-1})E$
where $E$ is the ejecta energy \cite{B02,B03c}. 
The initial soft flash rises sharply at the observer time
$\tobs\sim \Racc/\Gej^2 c$, which is before 10~s in most cases, so the 
rise may be difficult to catch with the current instruments.
However, the emission can last much longer after the rise, more than one 
minute, and then would be easily observed by {\it Swift}. Estimates
of the expected emission have been done in \cite{B02,Li03,Ku03}, however,
the problem requires a more accurate calculation that keeps track of 
the evolution of each shell in the blast material.

Blast waves in wind-type media ($n_0\propto R^{-2}$) have small $\Rdec$
\cite{LC03,KZ03,Wu03}, and the neglect of $e^\pm$ loading and 
preacceleration by $\gamma$-rays is inconsistent in this situation. 
The inclusion of these effects leads to
a very powerful prompt soft flash because most of the blast wave
energy dissipates at $R<\Rload$ \cite{B02}.
Its detection would be a clear signature of a massive progenitor.
On the other hand, the existing upper limits on the early optical 
emission in several GRBs exclude high-density winds in these bursts.

Optical flashes are also expected from the reverse shocks in the
GRB ejecta~\cite{MR93,SP99}. The reverse
shock with magnetic field $B$ comparable to that in the forward shock 
produces an optical flash like the one observed in GRB~990123 \cite{A99}.  
This can be used to probe the composition of the GRB ejecta as the reverse 
shock emission is sensitive to $B$. A magnetic field near or above 
equipartition with the fluid pressure is likely to
suppress the reverse shock component of the afterglow, and then only the 
$e^\pm$ emission from the forward shock contributes to the optical flash.
The forward-shock component does not depend on the nature of the ejecta 
and can help to determine the ambient density, magnetic field, and Lorentz 
factor of the blast wave.


\section{Neutron decay}

A significant fraction of baryons in the GRB ejecta are neutrons
\cite{DKK99,PWH03,B03a}. 
They are collisionally coupled to the ions when the fireball 
is accelerated by radiation pressure and develop a high $\Gamma_n=10^2-10^3$.
Then the neutrons decouple and coast with $\Gamma_n=const$. 

The $\beta$-decay depletes exponentially the neutron component outside
the mean-decay radius $\Rb\approx 10^{16}(\Gn/300)$~cm, which is 
comparable to the expected radius of the early afterglow.
However, the neutrons impact the blast wave 
at radii significantly larger than $\Rb$, even though their number is 
exponentially reduced at large radii. 

The front of survived neutrons 
overtakes the decelerating blast wave at some radius $R_*$ where the 
blast-wave Lorentz factor $\Gamma$ decreases below $\Gn$. Using the 
Blandford-McKee solution $\Gamma^2(R)=(17-4k)E/8\pi \rho_0 c^2R^3$ 
for adiabatic blast waves in a medium with density 
$\rho_0\propto R^{-k}$ we have
\be
 R_*^3=\frac{(17-4k)E}{8\pi \rho_0c^2\Gn^2}.
\ee
At $R>R_*$ the $\beta$-decay
takes place in the external medium {\it ahead} of the forward shock. 
The impact of this decay can be understood by comparing the energy of 
neutrons, $E_n=X_nE\exp(-R/\Rb)$ ($X_n$ is the initial neutron fraction 
of the explosion) with the ambient mass 
$mc^2=\frac{4\pi}{3-k}R^3\rho_0c^2=\frac{17-4k}{2(3-k)}(E/\Gamma^2)$ they 
interact with,
\be
   \frac{E_n}{mc^2}=\frac{2(3-k)}{17-4k}X_n\Gamma^2
                     \exp\left(-\frac{R}{\Rb}\right).
\ee
Just after $R_*$ this ratio can be as large as $\Gn^2$ depending on
$R_*/\Rb$. The decaying neutron front with $E_n>mc^2$ deposits
huge momentum and energy into the ambient medium, leaving behind a relativistic
trail. The exact parameters of this trail are found from energy and momentum
conservation applied to the collision of $\beta$-decay products with 
the ambient medium \cite{B03b}.

The ratio $E_n/mc^2$ becomes smaller than unity only after 
$\approx 10$ e-foldings of $\beta$-decay. Therefore, the impact of neutrons 
lasts until $\Rtrail\approx 10\Rb\approx 10^{17}$~cm, and 
one expects an observational effect if $R_*<\Rtrail$. For a 
homogeneous medium ($k=0$) this requires 
$n_0>0.1E_{52}(\Gamma_n/300)^{-5}$~cm$^{-3}$.
For a wind medium ($k=2$) $R_*<\Rtrail$ for all
plausible parameters of the wind if $\Gn\sim 10^2$ or higher.
Besides, the forward shock in a dense wind 
is likely to be slow from the very beginning (the reverse shock in the 
ejecta is relativistic). Then the neutrons can overtake the forward shock 
immediately, before the self-similar deceleration sets in.

The $\beta$-decay ahead of the shock transforms the cold static external
medium into a hot, dense, relativistically moving, and possibly magnetized,
material.
The ion ejecta follow the neutron front and 
drive a shock wave in the trail material. 
Dynamics and dissipation in the shock are discussed in \cite{B03b}.
Like the neutron-free shocks, it is difficult to calculate the expected 
synchrotron emission from first principles because the
electron acceleration and magnetic field evolution are poorly
understood. One can apply a phenomenological shock model with the customary
parameters $\epsilon_e$ and $\epsilon_B$ and fit the data with the model. 
This may enable an observational test for the $\beta$-decay.

Any neutron signature revealed in a GRB afterglow emission
would confirm that the ejected baryonic material has gone
through a hot high-density phase in the central engine.
Neutrons thus provide a unique link between the
physics of the central engine and the observed afterglow.
Numerical simulations of neutron-fed blast waves may help to identify
such signatures. One possibility, for instance, is an exponentially
decaying emission component.
Another possible signature is a spectral transition or a bump in the
afterglow light curve at $R\approx\Rtrail$~\cite{B03b}.
 
Absence of neutron signatures would indicate that the GRB ejecta are
dominated by magnetic fields. In such a low-density fireball, the neutrons
would decouple early with a modest Lorentz factor and decay quickly. 
Two-component ejecta with less collimated and less energetic
neutrons is possible in the MHD acceleration scenario~\cite{VPK03}.

\begin{theacknowledgments}
This research was supported by NASA grant NAG5-13382.
\end{theacknowledgments}

{}

\end{document}